\documentclass[conference]{IEEEtran}
\IEEEoverridecommandlockouts
\usepackage{cite}
\usepackage{amsmath,amssymb,amsfonts}
\usepackage{algorithmic}
\usepackage{graphicx}
\usepackage{textcomp}
\usepackage{xcolor}
\def\BibTeX{{\rm B\kern-.05em{\sc i\kern-.025em b}\kern-.08em
    T\kern-.1667em\lower.7ex\hbox{E}\kern-.125emX}}
\begin{document}

\title{Next-Generation High-Resolution Vector-Borne Disease Risk Assessment\\
}

\author{\IEEEauthorblockN{ Meysam Ghaffari}
\IEEEauthorblockA{\textit{Dept. of Computer Science} \\
\textit{Florida State University}\\
Tallahassee, FL, USA \\
ghaffari@cs.fsu.edu}
\and
\IEEEauthorblockN{Ashok Srinivasan}
\IEEEauthorblockA{\textit{Dept. of Computer Science} \\
\textit{University of West Florida}\\
Pensacola, FL, USA \\
asrinivasan@uwf.edu}
\and
\IEEEauthorblockN{Anuj Mubayi}
\IEEEauthorblockA{\textit{School of Human Evolution and Social Change} \\
\textit{Arizona State University}\\
Tempe, AZ, USA \\
anuj.mubayi@asu.edu}
\and
\IEEEauthorblockN{Xiuwen Liu}
\IEEEauthorblockA{\textit{Dept. of Computer Science} \\
\textit{Florida State University}\\
Tallahassee, FL, USA \\
liux@cs.fsu.edu}
\and
\IEEEauthorblockN{Krishnan Viswanathan}
\IEEEauthorblockA{
%\textit{dept. name of organization (of Aff.)} \\
\textit{Cambridge Systematics, Inc}\\
Tallahassee, FL, USA \\
kviswanathan@camsys.com}

}

\maketitle

\begin{abstract}
Vector-borne diseases cause more than 1 million deaths annually. Estimates of epidemic risk at high spatial resolutions can enable effective public health interventions. Our goal is to identify the risk of importation of such diseases into vulnerable cities at the granularity of neighborhoods. Conventional models cannot achieve such spatial resolution, especially in real-time. Besides, they lack real-time data on demographic heterogeneity, which is vital for accurate risk estimation. Social media, such as Twitter, promise data from which demographic and spatial information could be inferred in real-time. On the other hand, such data can be noisy and inaccurate. Our novel approach leverages Twitter data, using machine learning techniques at multiple spatial scales to overcome its limitations, to deliver results at the desired resolution. We validate our method against the Zika outbreak in Florida in 2016. Our main contribution lies in proposing a novel approach that uses machine learning on social media data to identify the risk of vector-borne disease importation at a sufficiently fine spatial resolution to permit effective intervention. This will lead to a new generation of epidemic risk assessment models, promising to transform public health by identifying specific locations for targeted intervention.

\end{abstract}


\section{Introduction}

Vector-borne diseases account for more than $17\%$ of infectious diseases, causing more than 1 million deaths annually from infections such as malaria, dengue fever, yellow fever, Japanese encephalitis, and schistosomiasis~\cite{WHO.2008}. Diseases once thought as controlled have experienced a resurgence. For example, there has been a 30-fold increase in Dengue compared with fifty years ago, and the World Health Organization has expressed concern about a potentially even higher rise in Zika~ \cite{world2016mosquito}. Over 2.17 billion people currently live in places suitable for the spread of Zika, which causes severe health problems to fetuses including microcephaly, congenital malformation of the brain, stillbirths, and damage to eyesight and hearing, leading to life-long disabilities~\cite{world2016mosquito,jones2018challenges}.  

Large scale spread of these diseases often happens by importation from a region experiencing a localized outbreak or a region where it is endemic. For example, the 2015 Zika outbreak in Brazil was related to an earlier introduction of the disease coinciding with a spurt in travel from Zika endemic areas \cite{faria2016zika}. The disease often subsequently spreads locally through local mosquitoes. For example, in the summer of 2009, local Dengue transmission was identified in Key West, FL after its importation through travel~\cite{Shin-2013}. {\em Our application focus is on places to which such diseases can be imported from endemic regions or regions experiencing an outbreak}. We will demonstrate our technique on the 2016 Zika outbreak in Florida, focusing on the importation from Puerto Rico. Most of the importation was from the Caribbean, with Puerto Rico being the largest single source, and we assume that the spatial pattern of importation from Puerto Rico reflects the broader pattern of importation from the Caribbean.

New developments in the cost of vector control and an increase in human travel limit the application of previous approaches to preventing the spread of several vector-borne diseases. Co-location of infected humans and vectors is a necessary condition for the introduction of the disease, and of infected vectors and susceptible humans for its spread \cite{little2017local,bogoch2016potential}. Eradication of Aedes mosquitoes in the 1960s removed one of the necessary conditions for the spread of diseases, such as Zika and Dengue. However, Aedes mosquitoes have resurfaced with an even greater spatial spread, accounting for locations with more than half the world's population. Insecticide resistance has led to the need for integrated vector control measures, which are too expensive to apply on a large scale \cite{world2016mosquito,velazquez2018vector}, leading to frequent disease outbreaks as shown in Table~\ref{tab:diseases}. 
Precise spatial identification of epidemic risk can enable cost-effective control. 

\begin{table}[htbp]
    \centering
    \caption{Recent incidences in Florida of vector-borne diseases  \cite{FDHPC,FloridaMorbidity,FloridaDeptHealth}}
   
    \begin{tabular}{|c |c |c| c| c| }
    
    \hline
        Disease & Year & Imported & Local & Source \\
       \hline
        Chikungunya	& 2014 & 475 & 12 & CDC\\
        \hline
         Dengue & 2009 & 10 & 66 & FDH \\
         \hline
         Dengue & 2013 & 70 & 28 & FDH\\
         \hline
         Zika & 2016 & 1016 & 256 & FDH\\ 

        \hline
    \end{tabular}

    \label{tab:diseases}
\end{table}

This requires estimating the risk, at a fine spatial scale, of infected and susceptible persons being present in places with an abundance of the vector~\cite{bouzid2016public}. Changes in travel patterns, for example, due to economic problems in Puerto Rico, make it challenging to acquire data on human movement and behavior with real-time information required in an emergency. Another challenge arises from conventional data for epidemic risk models not being available at the necessary spatial resolution. In addition, models of human movement have to account for demographic heterogeneity, because we need to estimate the travel patterns of a sub-population that has visited a disease-affected region. 

Social media data could provide real-time information, and also information from which spatial and demographic information can be inferred. However, it suffers from inaccuracies, noise, and paucity of geo-tagged data. We use a multi-scale approach to deal with these. We first identify the population flux of those who have visited Puerto Rico recently and then input it to a metapopulation epidemic model to identify risk at the county level. Computation of population flux leverages a low-resolution Natural Language Processing (NLP) technique for tweet location that uses tweet text content to deal with the paucity of geo-tagged tweets. We then identify specific vulnerable neighborhoods in high-risk counties using a computationally intensive deep-learning-based high-resolution Twitter user home-location prediction technique to deal with the noise and inaccuracies of geo-tagged tweet metadata. 

Our risk map at the level of a county accurately reflects the relative risks observed during the 2016 Zika epidemic. This resolution is novel in itself. More importantly, our risk map at the level of a neighborhood correctly identifies two of the three neighborhoods of the order of a square mile resolution each that CDC identified as high risk after observing the outbreak. Sources other than the Caribbean likely account for the one neighborhood that we missed\footnote{After we obtained our results, the Florida Department of Health confirmed in a personal email communication that the third neighborhood did not have a strong Puerto Rico connection to the outbreak, while the two that we identified did.}. Our results are a major breakthrough in epidemic modeling, and such a methodology could have helped mitigate the tragic impact of recent epidemics, such as those shown in Table~\ref{tab:zika}. 

The primary contribution of this paper lies in presenting an approach that can estimate epidemic risk at the neighborhood level using real-time data that accounts for demographic heterogeneity. This development can lead to a new generation of epidemic models that leverage social media data. Furthermore, Zika is spread by a mosquito genus, Aedes, which also spreads dengue, chikungunya, yellow fever, and West Nile Virus. Our methods can easily be extended to these diseases, and also to other geographical locations importing these diseases, including much of the Americas and Asia.

\begin{table}[htbp]
    \centering
     \caption{Recent incidences in Florida of Zika disease \cite{FDHPC,FloridaMorbidity,FloridaDeptHealth}}
   
    \begin{tabular}{|c |c |c| }
  \hline
        Year & Imported & Local \\
      \hline
        2016 & 1016 & 256 \\
        \hline
        2017 & 207 & 15 \\
        \hline
        2018 & 89 & 0 \\
        \hline
    \end{tabular}
    
    \label{tab:zika}
\end{table}

The rest of this paper is organized as follows. In Section~\ref{sec:background}, we present background material on epidemic models and the state of the practice in public health agencies. We present a high-level view of our proposed approach in Section~\ref{sec:approach} followed by data preparation in Section~\ref{sec:data}. Section~\ref{sec:county} discusses county-level prediction, while Section~\ref{sec:neighborhood} discusses neighborhood-level prediction. We summarize our conclusions and present directions for future work in Section~\ref{sec:conclusions}.

\section{Background}
\label{sec:background}

\subsection{Epidemic Models}

An epidemic model that predicts the number of infected persons in different geographical regions can be used to generate a Zika risk map. Agent based models and metapopulation models are popular for this purpose. Agent based models for epidemics, such as EpiSimdemics, use a synthetic population at the individual level and can address heterogeneity \cite{yeom2014overcoming,bhatele2017massively}. Meta-population models, such as in GLEaM~\cite{balcan2010modeling}, on the other hand, formulate the system as a directed graph, with vertices being geographic regions (called patches) and edges denoting the movement of people between patches with edge weight being a measure of the population flux.  

Agent based and metapopulation models have distinct strengths and weaknesses~\cite{ajelli2010comparing}. Agent based models provided a more detailed output, but require a large amount of good quality data as input and are also computationally intensive. Metapopulation models require less data and are computationally fast. This helps with new outbreaks, where epidemiological parameters are not known accurately.

Conventional models typically have the granularity at the level of countries or states. However, there has been much recent interest on the development of high-resolution models that use new sources of data, such as from social media, location based services, cell phones, air traffic, and land cover maps~\cite{brennan2013towards,zu2016dynamic,huff2016flirt}. Models for vector-borne diseases typically estimate vector characteristics, such as abundance, from models that use a variety of environmental variables including land cover, temperature, precipitation, etc. \cite{bogoch2016potential}. They then overlay this with human mobility data. However, such models for vector-borne diseases typically operate at the resolution of several counties. 

Recent works have attempted to obtain higher resolution. Little et al.~\cite{little2017local} have related vector characteristics with ecological factors in New York City to get risk estimation at the zip code level. However, this work does not consider human mobility in the introduction of the disease, which is known to play the primary role in disease spread. Given the large number of visitors to New York City, one could assume that the disease would be introduced into the city. However, the introduction may not be uniform in all zip codes, and so such a model is not accurate without accounting for demographic heterogeneity. In fact, their empirical results showed some discordant trends between disease prevalence and vector abundance. Bogoch et al.~\cite{bogoch2016potential} overlaid results of vector characteristics with air traffic data from affected countries. They determined vector-associated risk at 5 km$^2$ resolution. However, their model for human mobility did not account for heterogeneity in neighborhoods within a city to which people move after arriving in a city. Consequently, their validation was at the coarser scale of the country, with results showing that the 2013 Zika outbreak in Brazil coincided with a surge in the volume of airline travelers arriving from countries with Zika activity.

\subsection{Current State of Practice by Public Health Agencies}

Our research agenda was driven by extensive discussions with the Florida Department of Health (DoH) and Orange County (FL) Mosquito Control Division to assess the ground realities of public health practice where we could make an impact. The former is tasked with promoting public health in the state. However, vector surveillance and control are performed independently by county-level mosquito control districts. While the majority of public health agencies desire to use models, only a minority actually uses models due to their limitations for application in practice~\cite{Doms-2018}. 

Aedes mosquitoes are present throughout most of Florida. Consequently, human mobility plays the primary role in determining disease introduction and spread, as is true of most vulnerable regions~\cite{cosner2009effects}. Vector-borne diseases are typically introduced into Florida by travelers from Caribbean islands, and there is much fluctuation in this, especially exacerbated by extreme events such as hurricanes and economic turmoil. Established models lack real-time data on such movement. This is a particularly severe problem with Zika because $80\%$ of victims are asymptomatic and thus their mobility is not reduced, leading to wide spreading of the disease~\cite{kucharski2016transmission}. In addition, mosquito control districts are locally controlled and constrained by resources. Consequently, they cannot perform integrated control over a large region of the size of a city. They desire knowledge of locations at risk at a fine granularity so that both vector surveillance and mosquito control can be intensified where they are most relevant. This is a common situation~\cite{OrangeCounty,Surveillance}. 

\section{Our Approach}
\label{sec:approach}

\paragraph{Problem} Our goal is to identify locations at high-risk of a vector-borne disease outbreak, where the disease is imported from a different affected region and then transmitted locally by vectors. Identification of risk will be at the level of a neighborhood in a city, with spatial resolution of the order of a square mile. Our goal is not precise prediction of the number of morbid persons. Rather, it is on identification of specific locations that could be targeted for intensive public health intervention to limit the likelihood of a major epidemic.

\paragraph{Promises and Limitations of Twitter Data} Twitter provides access to three types of information from which location of a user may be inferred. This can be used to model human mobility and subsequently used in an epidemic model, especially considering that around $20\%$ of the US population have active Twitter accounts. One is the home location specified in a user profile. The other is a geo-tag that can be associated with each tweet. Finally, one may be able to infer location from the text contents of the tweet. Each comes with positive and negative aspects. Geo-tags provide high-resolution information on the location of a tweet. However, only around $1\%$ of tweets contain geo-tag. In addition, the location of a tweet does not necessarily indicate that the user spent considerable time there. The profile home location is available for a large fraction of users, and one can expect that users spend considerable time at home. On the other hand, this location is usually available only at a coarse granularity -- often at the level of a city or state -- and also does not capture mobility information. NLP techniques have been used to infer tweet location from its text contents, which can provide mobility information. However, their accuracy is relatively low and their spatial resolution is coarse.  

\paragraph{Basic Strategy} The basic novelty of our strategy lies in using Twitter data to determine, with high spatial resolution, the home locations of people who have visited a region already experiencing an outbreak of a disease. We leverage a new deep-learning algorithm that we recently developed to identify home locations of Twitter users with a reasonable number of geo-tagged tweets \cite{ghaffari2019high}. We had evaluated our algorithm on data from Chicago and were able to identify the home locations of over 80\% of such users with over 85\% accuracy at 100 meter resolution. (Home locations of the remaining 20\% of users were categorized as "unknown".) 

\paragraph{Challenges} There are multiple challenges to direct application of the above algorithm for vector-borne epidemic modeling. First, this algorithm is too computationally intensive to apply to all neighborhoods in a large region, such as Florida. Another is that only around 1\% of tweets contain geo-tags. Yet another is that such data can only provide an indication of the likelihood of importation of the disease, and not its local spread by simultaneously accounting for vector characteristics and human mobility. 

\paragraph{Multi-Scale Approach} We use a multi-scale approach to deal with the above challenges, as illustrated in Figure~\ref{fig:approach}. We first prepare a data sample that is likely to contain a much larger fraction of Twitter users who had visited Puerto Rico than the general population. We provide details in Section~\ref{sec:data}. We deal with the paucity of geo-tagged tweets by using a NLP algorithm that identifies location using tweet content at a coarse spatial resolution. We incorporate this into a metapopulation epidemic model to identify disease risk at the county level, accounting for local spread through mosquitoes and human mobility simultaneously. We explain this further in Section~\ref{sec:county}. We then focus on the most high-risk counties and apply our computationally intensive algorithm to identify the high-risk neighborhoods there, as explained in Section~\ref{sec:neighborhood}.

\begin{figure}
    \centering
    \includegraphics[scale=0.5]{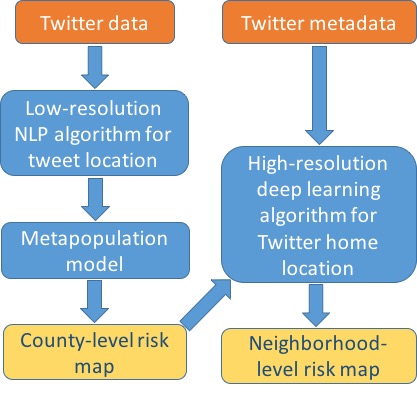}
    \caption{Overview of our approach}
    \label{fig:approach}
\end{figure}

\section{Data Preparation}
\label{sec:data}

We explain below our preparation of Twitter data and mention other data sets that we used. 

\paragraph{Florida-Puerto Rico (FL-PR) Twitter data.} We prepared our primary data, for prediction of mobility in Florida of people who had been to Puerto Rico, by extracting publicly available Twitter data using Twitter API. Our goal was to generate a sample that contained a large fraction of users who had been in both Florida and Puerto Rico. We extracted data through the following search procedure, whose effectiveness is based on the assumption that people with a strong connection to Puerto Rico are more likely to have visited it than the general population. It is analogous to the Response Driven Sampling procedure used in surveying populations that are difficult to reach~\cite{Salganik-2004}. We started with a seed user with profile home location in Puerto Rico. We extracted the list of that user's followers. We recursively applied the  same procedure to any follower whose profile home was in Puerto Rico or Florida. We repeated the same process with several seed users.  This yielded a list of over  500,000 users who either lived in Puerto Rico or Florida, or followed someone who lived there. We then downloaded tweets of these users. The Twitter API limits the number of tweets accessed per user to 3200, which limited our sample. We extracted this data over a span of 6 months, starting February 2018. Consequently, we obtained tweets in both 2016 and 2017. While the Zika epidemic was at its peak in Florida in 2016, we included the 2017 data so that our sample would be large. We assume that mobility patterns of our target sub-population would be relatively unchanged between the two years, and thus additional data from 2017 would be meaningful.

\paragraph{Florida morbidity data.} The Florida Department of Health (FDH) makes publicly available monthly morbidity data for various vector-borne diseases at the granularity of counties~\cite{FloridaDeptHealth}, including separate counts for imported and locally acquired cases. It also provides monthly data for the place of origin of imported cases, with the origins listed as countries or US territories. This data is provided only statewide, and not at the county level. Also, the data that we validate against is the total morbidity count, rather than those that were connected to importation from Puerto Rico, because the former data is publicly available. Centers for Disease Control (CDC) listed specific neighborhoods in Miami that were at risk of Zika infection during the 2016 outbreak, although they did not indicate the source of importation for the disease. 

\paragraph{Twitter-World-EX dataset.} This is a widely used dataset, which  was used to train a NLP algorithm~\cite{han2012geolocation} for tweet location prediction, while the actual use of the NLP algorithm is with the FL-PR Twitter dataset for county-level prediction. This dataset uses Twitter data from September 2011 till February 2012 for 29 million users from all over the world. 

\paragraph{Chicago dataset.} This dataset is based on geo-tagged Twitter metadata for tweets in Chicago between May 2014 and April 2015~\cite{kavak2018fine}. The Twitter metadata was processed by its authors to extract ten features for 1268 users with at least five geo-tagged tweets each. The home locations for all these users have been validated. We used this dataset to train our deep learning algorithm for high-resolution home location prediction.

\section{County Level Prediction}
\label{sec:county}

Our first goal is to identify those counties at the highest risk of disease importation and subsequent spread through local mosquitoes. We use a metapopulation epidemic model for this purpose. It requires population flux from Puerto Rico as input, which we estimate using Twitter data. We describe below our Twitter analysis and county-level risk prediction using the metapopulation model.

\subsection{Twitter Analysis for Population Flux}

The pigeo geo-tagging tool~\cite{rahimi2016pigeo} can be used to estimate tweet location and Twitter user location at coarse resolutions. We use it to determine tweet location. This tool divides the worlds into zones and assigns a tweet to a region using NLP. In particular, it uses a bag-of-unigrams model. Features vectors are based on binary term frequency, inverse document frequency, and $l_2$ normalization of samples. The model was trained using Twitter-World-EX dataset~\cite{han2012geolocation} and has an accuracy between 62\% to 67\% on global data at 161~km resolution~\cite{rahimi2016pigeo}. We used this model with the FL-PR twitter dataset to predict population flux from Puerto Rico to each Florida county as follows. We count the number of users with tweets in both Puerto Rico and Florida, and then calculate the fraction of such users with tweets in each county. Note that a user may visit Puerto Rico and then multiple Florida counties. That user will contribute to the estimate for each of these counties. We then multiply the fraction for each county by the air traffic from Puerto Rico to Florida to estimate the population flux from Puerto Rico to each Florida county.

\subsection{Epidemic Model}

We employ a metapopulation epidemic model to predict the number of infected persons in different counties, which we use as an estimate of risk. Metapopulation models formulate the system as a directed graph, with vertices being geographic regions, and edge weights $\alpha_{ij}$ denoting the population flux from vertex $i$ to $j$. We consider the vertices as the 67 counties of Florida plus Puerto Rico. While we include vector abundance in the model, we don't consider vector flux between counties because Aedes mosquitoes have a range of only 100-200m, which is small compared with the scale of a county. Variables used in the model are presented in Table~\ref{tab:Epidemic} and values of model parameters in Table~\ref{tab:EpidemicTuning}, based on values suggested in  literature~\cite{andraud2012dynamic,chikaki2009dengue,bearcroft1956zika,kucharski2016transmission,majumder2016estimating}. 

\begin{table}
\centering
\caption {Epidemic Model Parameters}
\begin{tabular}{ |c |c| } 
\hline
State & Variable Description \\ 
\hline
$S_{hi}$ & Susceptible Humans \\ 
\hline
$E_{hi}$ & Exposed Humans \\ 
\hline
$I_{hi}$ & Symptomatic Humans \\ 
\hline
$A_{hi}$ & Asymptomatic Humans \\ 
\hline
$R_{hi}$ & Recovered Humans \\ 
\hline
$S_{vi}$ & Susceptible Vectors \\ 
\hline
$I_{vi}$ & Infected Vectors \\ 
\hline
\end{tabular}

\label{tab:Epidemic} 
\end{table}

The equations below define the rate of change of model variables with time for each vertex $i$ with a total of $Z$ vertices. In our case, we consider the flux only from Puerto Rico. Consequently, the variable $j$ takes only one value, corresponding to Puerto Rico, while $i$ takes values only corresponding to the counties in Florida. The equations below are more general and can be used in situations covering fluxes between all vertices.

\begin{equation*}
S'_{hi} =-b* \beta_{vh}* \frac{I_{vi}*S_{hi}}{N_{hi}} + \Sigma_{j=1}^Z \alpha_{ji} * S_{hj} - \Sigma_{j=1}^Z \alpha_{ij} * S_{hi}
\end{equation*}
%\begin{equation*}
\begin{align*}
E'_{hi}=-b*\beta_{vh}*\frac{I_{vi}*S_{hi}}{N_{hi}} - \delta* E_{hi}+ \Sigma_{j=1}^Z \alpha_{ji}*E_{hj} - \nonumber\\
 - \Sigma_{j=1}^Z \alpha_{ij}*E_{hi} 
 \end{align*}
%\end{equation*}
\begin{equation*}
I'_{hi}= \delta*(1-\phi)*E_{hi} - \gamma*I_{hi} + \Sigma_{j=1}^Z \alpha_{ji}*I_{hj} - \Sigma_{j=1}^Z \alpha_{ij}*I_{hi} 
\end{equation*}
\begin{equation*}
A'_{hi}= \delta*\phi*E_{hi} - \gamma*A_{hi} + \Sigma_{j=1}^Z \alpha_{ji}*A_{hj} - \Sigma_{j=1}^Z \alpha_{ij}*A_{hi} 
\end{equation*}
\begin{equation*}
R'_{hi}= \gamma*(I_{hi}+A_{hi}) + \Sigma_{j=1}^Z \alpha_{ji}*R_{hj} - \Sigma_{j=1}^Z \alpha_{ij}*R_{hi} 
\end{equation*}
\begin{equation*}
S'_{vi}= -b* \beta_{hv}* \frac{(I_{hi}+A_{hi})*S_{vi}}{N_{hi}} - \mu * (S_{vi}-N_{vi})
\end{equation*}
\begin{equation*}
I'_{vi}= b* \beta_{hv}* \frac{(I_{hi}+A_{hi})*S_{vi}}{N_{hi}} - \mu *I_{vi}
\end{equation*}

Here, $N_{hi}$ is the human population in county $i$ and $\mu$ is the vector death rate. We solve the equations for steady state and estimate the risk for each county as $\frac{100}{365}I_{hi}$, where the multiplier assumes an active mosquito season of 100 days a year. Note that the epidemiological parameters for Zika are not well understood. Consequently, the predictions above should be interpreted as estimates of relative risk between counties, rather than as actual estimates of morbidity.  

\begin{table}[htbp]
\centering
\caption {Epidemic Model Tuning Parameters}
\begin{tabular}{ |c |p{2.5cm}| p{1.5cm} |} 
\hline
Parameter & Description & Value \\
\hline
b & Bitting Rate & 0.5 per day  \\ 
\hline
$\beta_{hv}$ & Human to vector infection probability & 0.5 per day\\ 
\hline
$\beta_{vh}$ & Vector to human infection probability & 0.4 per day  \\ 
\hline
$\frac{1}{\delta}$ & Intrinsic Incubation period & 5 days \\ 
\hline
$\gamma$ & Recovery rate &  0.25 per day  \\ % \cite{gao2016prevention}
\hline
$\phi$ & Proportion of asymptomatic infections & 0.18   \\ % \cite{gao2016prevention}
\hline
\end{tabular}

 \label{tab:EpidemicTuning} 
\end{table}

\subsection{County Level Risk Assessment}

\begin{figure}[htbp]
    \centering
    \includegraphics[scale=0.4]{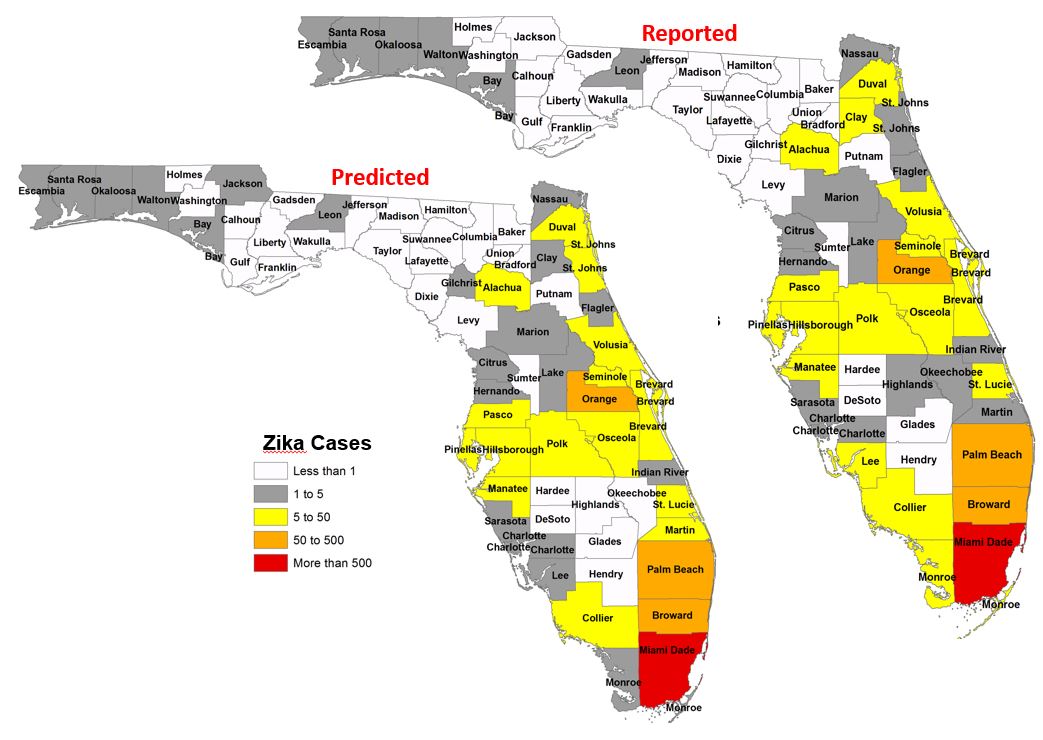}
    \caption{The generated County level risk map}
    \label{fig:zika-incidence}
\end{figure}

Figure~\ref{fig:zika-incidence} compares the relative risk in different counties, both as estimated by our method and based on actual prevalence in 2016. We can see that our model matches the relative risk observed in practice in most of the counties, except for slight differences in some low-risk counties.  It correctly identifies the high and medium-risk counties and identifies Miami-Dade as the county with the highest risk. 

\begin{table}[htbp]
    \centering
     \caption{Number of Reported Zika cases in 2016 \cite{FloridaMorbidity}}
    \begin{tabular}{|c| c| c| c| }
   \hline
         County & Travel & Local & Total  \\
        \hline
         Miami Dade & 350 & 287 & 681 \\
         \hline
         Broward & 182 & 1 & 183 \\
         \hline
         Orange & 176 & 0 & 167 \\
         \hline
         Hillsborough & 46 & 0 & 46 \\
         \hline
          Lee & 15 & 0 & 15  \\
         \hline
         Alachua & 12 & 0 & 12 \\
         \hline
         Duval & 11 & 0 & 12 \\
         \hline
         Sarasota & 5 & 0 & 5 \\
         \hline
         Volusia & 12 & 0 & 2 \\
         \hline
         Leon & 2 & 0 & 2  \\
         \hline
    \end{tabular}
    
    \label{tab:zika-incidence}
\end{table}

\section{Neighborhood Level Prediction}
\label{sec:neighborhood}

The analysis of Section~\ref{sec:county} shows that the estimated (and actual) Zika incidence in Miami-Dade county is more than that of all other counties combined. Consequently, we focus our high-resolution search solely on this county. Metapopulation models cannot be used at this resolution because infected population levels are too small. Consequently, population flux is not required. Instead, we focus on locations of persons who have been in both Miami and Puerto Rico. We will consider Puerto Rico residents who have visited Miami and Miami residents who have visited Puerto Rico. 

\subsection{Puerto Rico Residents Who Visited Miami}
These Twitter users can be identified through their home profile location indicating Puerto Rico with tweets having geo-tags in Miami. We determine neighborhoods visited from the geo-tag. One deficiency of this result is that locations visited may not indicate the time spent there, and consequently the risk. For example, a Puerto Rico resident may visit a relative in Florida, but tweet predominantly from tourist attractions.

Table \ref{tab:MimamiNeighborhoods} shows the percentage of people  who lived in Puerto Rico and visited Miami.

\begin{table}[htbp]
    \centering
    \caption{Percentage of visited locations in different Miami neighborhoods for users who live in Puerto Rico}
    \begin{tabular}{|c |c|}
    \hline
        Neighborhood & Percentage visits\\
        \hline
         Miami International Airport & 	$16.9\%$ \\
         \hline
         Marlin Parks & $ 14.2\%$ \\
         \hline
         Wynwood & $14\%$ \\
         \hline
         InterContinental & $13.5\%$ \\
         \hline
         Miami Beach & $10\%$ \\
         \hline
    \end{tabular}
    
    \label{tab:MimamiNeighborhoods}
\end{table}

\subsection{Miami Residents Who Visited Puerto Rico}

We now consider the risk of a Miami resident who may have got infected on a visit to Puerto Rico and subsequently returned to Miami infected. Residents spend considerable time around their home neighborhood, and so neighborhoods with a large number of such people could be at risk. Twitter profile home location does not normally provide information at the granularity of a neighborhood. We, therefore, use our high-resolution deep learning algorithm for home-location prediction based on metadata of geo-tagged tweets to resolve this problem. Our algorithm is designed to predict home location to within 100m. 

Our algorithm uses that same features as~\cite{kavak2018fine}, although our algorithm is very different from it. We summarize important aspects below. First, we apply DBSCAN to cluster tweets of a user that are within a range of 100 meters of each other. We create a record for each cluster that includes ten features related to geo-tag location,  time, sequence of activities, and activities of other users at location~\cite{kavak2018fine}. For example, if a user has several tweets with geo-tag within the range of 100 meters from each other at 7PM-12AM, it is considered as the end of day activity. We trained our algorithm on a dataset from Chicago~\cite{kavak2018fine}. We then created a similar dataset for Twitter users with geo-tags in Puerto Rico and Miami and used our algorithm to determine home locations of users, and noted the home location that were in a neighborhood of Miami. Figure~\ref{fig:approach} summarizes our algorithm.

\begin{figure}
    \centering
    \includegraphics[scale=0.5]{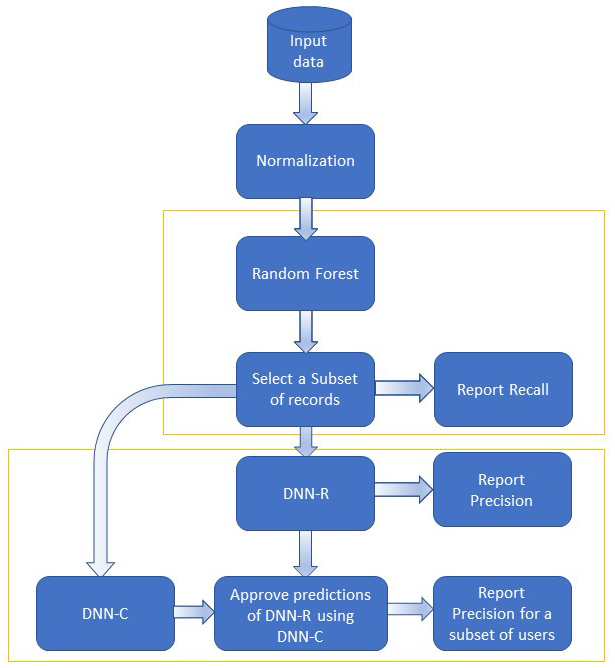}
    \caption{Overview of home location prediction method using dynamic structure and deep neural network [from~\cite{ghaffari2019high}].}
    \label{fig:approach}
\end{figure}

Most of the tweets are not from the home location, leading to a very biased data set. We use a dynamic structure that uses a fast algorithmic component to quickly prune a portion of false records with high recall and then use more complex methods to precisely detect the users' home location among the available tweet locations. 

\paragraph{Random forest} The first phase is a random forest. It consists of a large number of trees, each of which predicts whether a record is the true home of that user. We prune a record if none of the trees categorizes it as the true home. 

\paragraph{Deep Neural Network Model for Regression (DNN-R)} We use a sequential fully connected DNN with five dense layers, and an additional four dropout layers to prevent overfitting, using a Stochastic Gradient Descent optimizer. For each record from the pruned set, it predicts the probability of that record being the home and outputs the highest probability record for each user as the candidate home. 

\paragraph{Deep Neural Network Model for Classification (DNN-C)} This is a  sequential fully connected DNN with five dense layers and four dropout layers, using RMSProp optimizer. It is input the set of pruned records and the candidate home record from DNN-C. It categorizes this candidate record as either the true home or rejects it. Users for whom the candidate home is rejected are considered to have an unknown home.

\subsection{Results}

We present below the most popular Miami neighborhoods for (a) Puerto Rico residents and (b) Miami residents who visited Puerto Rico. Each result has its deficiencies. With the former, geo-tag location may not indicate time spent. With the latter, the home location is inferred through an algorithm, rather than directly being available from the data. In order to produce robust results, we identify a neighborhood at high risk if it is indicated as high risk according to both criteria. On the other hand, public health officials may prefer to be on the safe side and target high-risk neighborhoods in either category for intensive intervention.

\textbf{Analysis of users with trips to Puerto Rico:}
Table \ref{tab:Percentage} shows the percentage of people  who visited both Miami and Puerto Rico in 5 detected neighborhoods reported in section 6.1.

\begin{table}[htbp]
    \centering
    \caption{Percentage of home locations of users in different Miami neighborhoods}
    \begin{tabular}{|c |c|}
    \hline
        Neighborhood & Percentage of detected home locations\\
        \hline
        Down Town & $25\%$ \\
         \hline
         Miami Beach & 	$20\%$ \\
         \hline
         Wynwood & $ 10\%$ \\
         \hline
         Miami Airport & $10\%$ \\
         \hline
         Allapattah & $10\%$ \\
         \hline
         %InterContinental & $1.6\%$ \\
         
    \end{tabular}
    
    \label{tab:Percentage}
\end{table}

Our results show that the following neighborhoods were identified as high risk through both criteria: Miami Beach, Wynwood, and Miami International Airport. The first two were also identified by CDC as high-risk "red zones" well into the outbreak. On the other hand, it incorrectly identifies the airport neighborhood as at high risk. Twitter geo-tags show the Miami residents tweeted likely within the airport area. Given the context of the airport, public health officials would probably understand that the airport is not likely to be a place where users spend much time exposed to vectors. Given our algorithm's 100m resolution, we were also able to examine the specific location of the purported home of Miami residents in the airport neighborhood. These were not in the airport but in locations in proximity to the airport. We need to examine this further to determine if our algorithm erred on the home locations of these users, or if local ecological factors limited Zika spread. CDC had identified Little River as another high-risk zone, which our algorithm did not capture. Florida Department of Health researchers confirmed through email that this neighborhood did not have a strong Caribbean connection, while the two that we identified did.

\section{Conclusions}
\label{sec:conclusions}

We show that we can identify locations at risk of importation of Zika at high spatial resolution using Twitter data. Our risk map at the county level was very accurate, and we also correctly identified the two high-risk Miami neighborhoods associated with the importation of the disease from the Caribbean. This is an unprecedented resolution for vector-borne disease risk estimation using data sources that can deliver real-time information while providing data to account for demographic heterogeneity. As mentioned in Section 2, the current state of the art cannot predict the risk of vector-borne diseases at the fine spatial granularity that we obtain. Furthermore, our machine learning models were trained on unrelated data. Thus, our results promise to be generalizable.

We expect this work to lead to next-generation high-resolution models that have an impact on public health. Our algorithm can directly be transferred to diseases such as Dengue and Chikungunya that are spread by the same Aedes mosquitoes.

In future work, we will examine the reason for reducing false positives (Miami International Airport in this instance). We will determine if the method can be made more effective by including other data sources, such as  location based services and Yelp data.

\bibliographystyle{IEEEtran}


\vspace{12pt}

\end{document}